# Coexistence of superconductivity and incoherence in quasi 1D chains

R. Khasanov†*, T. Kondo†, J. Schmalian†, S. M. Kazakov‡, N. D. Zhigadlo‡, J. Karpinski‡, H. M. Fretwell†, H. Keller*, J. Mesot¶ and Adam Kaminski†

†Ames Laboratory and Department of Physics and Astronomy, Iowa State University, Ames, IA 50014, USA

*Physik-Institut der Universität Zürich, CH-8057 Zürich, Switzerland

‡ Laboratory for Solid State Physics ETH Zürich, CH-8093 Zürich, Switzerland

¶Laboratory for Neutron Scattering, ETH Zürich and Paul Scherrer Institute, 5232 Villingen PSI, Switzerland

**The dimensionality of a correlated many-body system has a large impact on its electronic properties. When electrons are confined to one-dimensional chains of atoms their behavior is very different than in higher dimensional systems because they become strongly correlated, even in the case of vanishingly small interactions. The chains consisting of copper and oxygen atoms are particularly interesting, because the CuO orbitals are constituents of all known high temperature superconductors. Most of previous spectroscopic studies of CuO chain systems indicated insulating behavior[23-25]. Here we report the discovery of momentum dependent superconducting gap and hump-peak-dip structure in the spectra of the CuO chains. We demonstrate that superconductivity in the chains arises due to proximity effects and the peculiar momentum dependence of the superconducting gap shows how three dimensional coherence emerges in a layered superconductor. The presence of the hump-dip-peak structure in the spectra of the CuO chains is very unexpected as it was thought to only occur in the $CuO_2$ planes and is frequently considered to be a signature of the d-wave pairing and the pairing boson itself.**

We used Angle Resolved Photoemission Spectroscopy (ARPES) with a very small (50 μm) synchrotron beam and were able to directly identify two different types of termination on the surface for the cleaved $YBa_2Cu_4O_8$ samples (experimental details are described in Methods section). The first type (plane domains) displayed an electronic structure consistent with a quasi two dimensional (2D) electron system and well resolved bilayer splitting indicating that for these portions, cleaving exposed a BaO layer with $CuO_2$ planes directly below. The second type of termination (chain domains) was far more interesting as the signal was consistent with quasi-one-dimensional (1D) electronic structure. In this case, cleaving exposed the CuO chains at the surface. The very small escape depth of the photoelectrons allowed us to clearly separate the two domains by simply selecting an appropriate location on the surface of the sample. We have repeated our measurements on several samples and always found a few domains (~ 50-100 μm in size) that were terminated with either CuO chains or $BaO/CuO_2$ planes. For the remainder of this letter we will focus our attention on the signal originating from the CuO chains. The diagram of the experimental situation is shown in figure 1a, where $k_x$ and $k_y$ denote the momenta of electrons parallel and perpendicular to the direction of the chains, respectively. In figure 1b we plot a typical intensity map of the photoelectrons as a function of $k_x$ and binding energy (referenced to the chemical potential) obtained at the zone boundary ($k_y=\pi/a$) deep in the superconducting state at T=40K. The high intensity regions (red, yellow and green) mark the dispersion of the bands. In figure 1c we plot energy distribution curves (EDCs) corresponding to the data in the right half of the color plot (1b). Two main features are easily identified in these data and both are quasi one-dimensional. The most prominent feature is a strongly dispersive conduction band seen as the upper U-shaped area of high intensity in the color plot and a series of peaks close to the chemical potential in the EDCs. As this band approaches the chemical potential the peaks become very sharp, consistent with the presence of long-lived quasiparticles. Another striking property of the conduction band is the renormalization of the dispersion close to the Fermi energy (rapid change of velocity visible as a "kink" in the color plot and labeled with an arrow). This feature together with the hump-dip-peak structure in the EDC lineshape is very similar to what is observed in the quasi 2D $CuO_2$ planes of $Bi_2Sr_2CaCu_2O_{8+\partial}$ [12-14]. These observations are normally interpreted as being due to a very strong interaction with a bosonic mode. Recent inelastic neutron scattering experiments on untwinned $YBa_2Cu_3O_{7-\delta}$ have revealed a strong in-plane anisotropy of the spin excitations [15], which has been interpreted both in terms of evidence for stripe formations [16] or resulting from orthorhombicity effects within more conventional Fermi-liquid approaches [17]. Our present ARPES data indicate that the coupling between the chains and planes cannot be neglected and it must be considered in any model aimed at reproducing the measured dynamical spin susceptibility of YBCO compounds. The second feature present in the data is an insulating band at higher binding energy that has an energy gap of about 200 meV. At present the origin of the second band and the nature of its gap are not clear. It is unlikely to be due to bilayer splitting. The two bands are separated by ~500 meV which is an order of magnitude higher than the theoretically predicted value of bilayer splitting (36 meV) in the double CuO chain system [18]. Also, if bilayer splitting gives rise to the two bands, any common mechanism leading to an energy gap (e. g. a charge density wave or a Mott Hubbard gap) would affect both bands in the same way. Here only one of the bands exhibits an energy gap. An alternative, more plausible explanation is that the two chains have significantly different carrier concentrations due to the different chemical surroundings at the surface. The conduction band thus originates from chains further from the surface and its carrier concentration and electronic properties are closer to that of the bulk where the chains are known to be metallic [19]. The gapped band is therefore due to chains at the surface that are better isolated from the rest of the system.

Figure 2a shows the intensity of the photoelectrons at the chemical potential plotted as a function of $k_x$ and $k_y$. The areas of high photoelectron intensity (green, yellow and red) correspond to the Fermi surface, which shows remarkable quasi 1D character. Below in figure 2bcde we plot the data along several cuts indicated by the solid lines in (a). The overall high-energy dispersion of the conduction band is very similar for all values of $k_y$ momentum (except for a slight variation in the energy of the bottom of the band), again consistent with its quasi 1D character.

We now examine the low energy features of the conduction band in more detail. In figure 3abcd we plot data along several cuts through the Fermi surface. In panels 3efgh we plot the corresponding EDC spectra. The remarkable thing about this data is that it can be classified into three categories. For large $k_y$ (d,h) the data is coherent (sharp quasiparticles are clearly visible at $k_F$) and renormalization effects are moderate, i.e. the low binding energy velocity above ~ -55 meV (obtained from the rate at which the peak disperses) is moderately smaller than the one at high binding energy below 55meV. As $k_y$ decreases, the low energy dispersion becomes much smaller leading to a far bigger renormalization of the velocity ("kink"), and the state remains coherent as evident by the presence of sharp quasiparticle peaks (panels 3bf and 3cg). EDC spectra in this momentum range display very pronounced hump-dip-peak feature that is frequently observed in superconducting two dimensional $CuO_2$ planes. In panel 3kl we plot EDC at two momentum values in below and above the bulk $T_c$. Just like in case of spectra from $CuO_2$ planes, this feature exists only in the superconducting state. "Hump-dip-peak" structure along with "kink" in dispersion are considered spectral hallmark of high temperature superconductivity and is frequently attributed to a pairing boson. Fact that we now see those features in 1D CuO chains is very unexpected and may potentially lead to better understanding of their origin and relation to the mechanism of high temperature superconductivity. Finally, for small values of $k_y$ the coherent peak vanishes (3ae). Such behavior is very unusual, because the sample is in the superconducting state and one would expect all states at different $k_y$ to be at least coherent. This is in sharp contrast to the behavior of two-dimensional nodal states (such as in $Bi_2Sr_2CaCu_2O_{8+\partial}$) that albeit gapless, are coherent at low temperature [20]. The above behavior is again illustrated in figure 3i, where we plot EDC spectra at the Fermi momentum along the quasi 1D Fermi surface. Spectra close to the zone boundary ($k_y=\pi/a$) display a strong coherent peak, which vanishes around $k_y=0.3$ $\pi/a$ and then reappears at $k_y=0$. In figure 3j we plot the estimated weight of the coherent peak as a function of momentum $k_y$.

Further unusual behavior is seen in the superconducting gap. By analyzing data for a range of $k_y$ we find that only a handful of states develop a superconducting gap. The presence of the superconducting gap can be determined via symmetrization [21]. In figure 4 we show several such plots for a few cuts indicated in the Fermi surface model of fig. 4g. If a gap is absent in the spectra, then one observes two peaks on the occupied side at positive and negative energy that merge at $k_F$ into a single peak which quickly vanishes beyond $k_F$, such as in panels 4aef. The symmetrized spectra that have a superconducting gap look very different (4bcd). As before, the two peaks on the occupied side approach the chemical potential. However, this time they do not merge into a single peak at $k_F$ but remain separated well beyond $k_F$ due to particle hole mixing. We have confirmed these conclusions by comparing spectra at the Fermi momenta ($k_F$) with spectra obtained from a non-superconducting reference. The magnitude of the gap at $k_F$ has been measured for different $k_y$ and is shown in figure 4h. Close to normal emission ($k_y = 0$) there is no gap. For momenta larger than $k_y = \sim 0.5$ $\pi/a$ a small superconducting gap opens. The magnitude of the gap decreases as $k_y$ increases further and it closes again beyond $k_y = 0.8$ $\pi/a$.

Our results demonstrate that the superconductivity in quasi 1D chains has a non-trivial character and unlike in higher dimensions it affects only some of the electronic states.

The most natural explanation for this behavior is chain induced superconductivity via a proximity effect. The largest superconducting gap in the chains and the strongest quasi-particle renormalization occur for those chain states where the chain and planar Fermi surfaces cross. This way gapped planar states with the same momentum ($k_x$, $k_y$) are close to the chain Fermi energy, making the proximity coupling most effective. This is remarkable as it requires momentum conservation for the Cooper-pair motion between the chains and planes, while the corresponding single particle motion was frequently modeled assuming fully incoherent chain-plane coupling [22]. Our data also demonstrate that single particle coherence in the chains behaves very differently from Cooper pair proximity coupling. Chain states close to $k_y=\pi/a$ and close to the plane Fermi surface become coherent but only those close to the chain-plane Fermi surface crossing form a superconducting gap (maximum gap for the chain states is marked by a circle in Figure 4g). Again this can be explained by assuming that the motion of the Cooper pairs from the planes to chains conserves momentum while the momentum along the chains is not conserved for single particle tunneling.

At higher energies and temperatures one can expect Luttinger liquid behavior of the quasi one-dimensional chains. Theoretical evidence for the formation of a Luttinger liquid in the double chains of $YBa_2Cu_4O_8$, and against a spin-gapped state, was given in Ref. [18]. This would be fully consistent with the observation of Luttinger liquid behavior in $PrBa_2Cu_4O_{8+x}$ and $SrCuO_2$ seen in optical and photoemission experiments [23-25]. The latter compounds however are insulators and not superconductors. For a Luttinger liquid one expects a holon branch of the spectral function $\propto (\omega - v_c k)^{(\alpha-1)/2}$ with charge velocity $v_c > v_F^0$ and spinon branch $\propto (\omega - v_s k)^{\alpha-1/2}$ with velocity $v_s < v_F^0$. Here $v_F^0 \approx ta$ is the Fermi velocity of the chains in the absence of correlations and $\alpha$ is the non-universal Luttinger liquid exponent that is directly related to the interaction strength. Similar to Ref. [23-25] one can argue that the high-energy part of the conducting band shown in Fig.1b is the holon branch of the spin-charge separated Luttinger liquid. The features in the low energy part of the spectrum can either be due to the spinon branch, if $\alpha < \frac{1}{2}$, or due to the onset of coherent states caused by the coupling, $t_\perp$, to the planes [26]. The latter occurs for energies below $t(t_\perp/t)^{1/\phi}$ with $\phi = 1 - \frac{\alpha}{2}$ i.e. if $\alpha < 2$. Based on the $k_y$-dependence of the low energy spectrum, we conclude that the low energy feature is most likely not a spinon branch, which should occur for all $k_y$ equally, except for the unlikely case that $\alpha$ crosses $\frac{1}{2}$ sharply as a function of $k_y$. We then obtain $\frac{1}{2} < \alpha < 2$ not unusual for a Luttinger liquid with not too short ranged repulsive interactions. The absence of coherency close to $k_y = \pm 0.15$ π/a is therefore most interesting, as it demonstrates that those states remain uncoupled to the planar quasiparticles, i.e. remain Luttinger liquids while a dimensional crossover, similar to the one observed earlier for 2D and 3D systems [27], occurs for other $k_y$.

Our observations of a momentum dependent onset of coherency of chain states should have important implications for the onset of three-dimensional coherency of the system. Coherent chain-plane coupling is needed to achieve three-dimensional coherency of the $YBa_2Cu_4O_8$ structure. Coherent c-axis transport with $\frac{d\rho_c}{dT} > 0$ sets in below 200K [3]. On the other hand, the T-dependence of the plane and chain resistivity are qualitatively different down to $T_c$ and the electronic states in the $CuO_2$-planes and the CuO-chains behave fundamentally different, down to the lowest temperatures: the temperature dependence of the Cu-spin lattice relaxation rate $^{63}T_1^{-1}$ clearly demonstrates the absence of a spin gap in the chains while planes are governed by the pseudogap [18]. We believe that the selective emergence of chain coherency identified in this paper may offer the clue for the understanding of these interesting and seemingly contradictory observations.

## METHODS

The samples of $YBa_2Cu_4O_8$ ($T_c$=80K) were grown using a high-pressure flux method [4,5] and were cleaved in-situ in the ARPES system at low temperature and pressure (better than $5x10^{-11}$ Tr). The samples were oriented to make the direction of the component of the electric vector parallel to the cleavage plane also parallel to the chain direction [9]. The energy spectra of the photoelectrons were recorded as a function of emission angle using a high-resolution (0.1 deg angular and 12-16 meV energy) electron analyzer SES2002 and the PGM beamline at the Synchrotron Radiation Center, University of Wisconsin at Madison. The data in this letter was obtained using 22 eV and 33 eV photons. Previous scanning tunneling microscopy (STM) [6,7] of a related compound - $YBa_2Cu_3O_{7-\delta}$ revealed that the preferred cleavage plane lies between the BaO layer and CuO chains, consistent with early angle resolved photoemission spectroscopy studies of both $YBa_2Cu_3O_{7-\delta}$ [8-10] and $YBa_2Cu_4O_8$ [11].

[1] Bucher, B., Steiner, P., Karpinski, J., Kaldis, E., and Wachter, P. Influence of the spin gap on the normal state transport in $YBa_2Cu_4O_8$. *Phys. Rev. Lett*. **70,** 2012–2015(1993).

[2] Hussey, N. E., Nozawa, K., Takagi, H., Adachi, S. and Tanabe, K. Anisotropic resistivity of $YBa_2Cu_4O_8$: Incoherent-to-metallic crossover in the out-of-plane transport*, Phys. Rev. B* **56**, R11423–R11426 (1997).

[3] Hussey, N. E. *et al*. Magnetic Field Induced Dimensional Crossover in the Normal State of $YBa_2Cu_4O_8$ *Phys. Rev. Lett*. **80**, 2909–2912 (1998).

[4] Karpinski, J., Kaldis, E., Jilek, E., Rusiecki, S. and Bucher, B. Bulk synthesis of the 81-K superconductor at high-oxygen pressure. *Nature* **336**, 660-662 (1988)

[5] Karpinski J. *et al*. High-pressure synthesis, crystal growth, phase diagrams, structural and magnetic properties of Y(2)Ba(4)Cu(n)O2(2n+x), HgBa(2)Ca(n-1)C(n)O(2n+2+∂) and quasi-one-dimensional cuprates. *Supercond. Sci. Technol*. **12,** R153-R181 (1999)

[6] Edwards, H. L., Markert, J. T. and de Lozanne, A. L. Energy gap and surface structure of $YBa_2Cu_3O_{7-x}$ probed by scanning tunneling microscopy. *Phys. Rev. Lett*. **69**, 2967 (1992)

[7] Derro, D. J. *et al*., Nanoscale one-dimensional scattering resonances in the CuO chains *of $YBa_2Cu_3O_{6+x}$* *Phys. Rev. Lett*. **88**, 097002 (2002)

[8] Cmpuzano, J. C. *et al*. Fermi surfaces of $YBa_2Cu_3O_{6.9}$ as seen by angle resolved photoemission. *Phys. Rev. Lett.* **64**, 2308-2311 (1990)

[9] Lu, D. H. *et al.* Supercondcuting gap and strong in-plane anisotropy in untwinned $YBa_2Cu_3O_{7-x}$. *Phys. Rev. Lett*. **86**, 4370-4373 (2001)

[10] Schabel, M. C. *et al*. Angle resolved photoemission on untwined $YBa_2Cu_3O_{6.95}$ I. electronic structure and dispersion relations of surface and bulk bands. *Phys. Rev. B* **57**, 6090-6106 (1998)

[11] Gofron, K. *et al.* Observation of an "Extended" van Hove singularity in $YBa_2Cu_4O_8$ by ultrahigh energy resolution angle resolved photoemission. *Phys. Rev. Lett.* **73**, 3302-3305 (1994)

[12] Valla, T. *et al*. Evidence for quantum critical behavior in the optimally doped cuprate $Bi_2Sr_2CaCu_2O_{8+\partial}$. *Science* **285**, 2110 (1999).

[13] Bogdanov, P. V. *et al.* Evidence for an energy scale for quasiparticle dispersion in $Bi_2Sr_2CaCu_2O_8$. *Phys. Rev. Lett.* **85**, 2581-2584 (2000).

[14] Kaminski, A. *et al.* Renormalization of spectral lineshape and dispersion below Tc in $Bi_2Sr_2CaCu_2O_{8+\partial}$. *Phys. Rev. Lett.* **86**, 1070-1073 (2001).

[15] Hinkov, V. *et al.* LTwo-dimensional geometry of spin excitationsin the high-transition-teperature superconductor $YBa_2Cu_3O_{6+x}$. *Nature* **430**, 650-653 (2004)

[16] Seibold, G. and Lorenzana, J. Doping dependence of spin excitations in the stripe phase of high-$T_c$ superconductors. *Phys. Rev. B* 73, 144515 (2006)

[17] Eremin, I. and Manske, D. Fermi-liquid-based theory for the in-plane magnetic anisotropy in untwined high-$T_c$ superconductors. *Phys. Rev. Lett.* 94, 067006 (2005)

[18] Muller, T. F. A. and Rice, T. M. Luttinger-liquid state of the zigzag double chain, *Phys. Rev. B* **60**, 1611-1616 (1999).

[19] Lee, Y.-S., Segawa, K., Ando, Y. and Basov D. N. Coherence and superconductivity in coupled one-dimensional chains: a case study of $YBa_2Cu_3O_y$.

[20] Kaminski, A. *et al.* Quasiparticles in the superconducting state of $Bi_2Sr_2CaCu_2O_{8+\partial}$. *Phys. Rev. Lett.* **84**, 1788-1791 (1999)

[21] Norman, M. R. *et al.* Destruction of the Fermi surface in underdoped high-$T_c$ superconductors. *Nature* **392**, 157-160 (1998)

[22] Morr, D. K. and Balatsky, A. V. Proximty effects and quantum dissipation in the chains of $YBa_2Cu_3O_{6+x}$ *Phys. Rev. Lett.* **87** 247002 (2001)

[23] Mizokawa, T. *et al.* Angle-resolved photoemission study of insulating and metallic Cu-O chains in $PrBa_2Cu_3O_7$ and $PrBa_2Cu_4O_8$, *Phys. Rev. Lett.* **85**, 4779-4782 (2000)

[24] Mizokawa, T. *et al.* Angle-resolved photoemission study of Zn-doped $PrBa_2Cu_4O_8$: possible observation of single-particle spectral function for a Tomonaga-Luttinger liquid. *Phys. Rev. B.* **65**, 193101 (2002)

[25] Kim, B. J. *et al.* Distinct spinon and holon dispersions in photoemission spectral functions from one-dimensional $SrCO_2$. *Nature physics* **2**, 397-401 (2006).

[26] Carlson, E. W. *et al.* Dimensional crossover in quasi-one-dimensional and high-$T_c$ superconductors. *Phys. Rev. B.* **62**, 3422-3436 (2000)

[27] Valla, T. *et al.* Coherence-incoherence and dimensional crossover in layered strongly correlated metals. *Nature* **417**, 627-630 (2002).

**Acknowledgements** We acknowledge enlightening discussions with M. Vojta, M. Eschrig and S. A. Kivelson. This work was supported by the Director for Energy Research, Office of Basic Energy Science, U. S. Department of Energy. R. K. was supported by Swiss National Science Foundation and K. Alex Müller Foundation. Ames Laboratory is operated for U. S. DOE by Iowa State University under contract No. W-7405-Eng-82. The Synchrotron Radiation Center is supported by NSF under grant No. DMR-9212658.

**Competing interests statement** The authors declare that they have no competing financial interests.

**Correspondence** and requests for materials should be addressed to A. Kaminski (email: kaminski@ameslab.gov).

**Figure 1** Experimental arrangement and typical data obtained from chain domains using 22 eV photons at T=40K. **a**, Structure on the surface of the sample in a chain domain. $k_x$ and $k_y$ are parallel and perpendicular to the 1D chains, respectively. Normal emission is at $k_x=k_y=0$. Typical intensity map **b** and EDCs **c** taken along the chains at $k_y=\pi/a$.

**Figure 2** Fermi surface and band dispersion in quasi 1D chains. **a**, Intensity of the photoelectrons obtained by integrating ARPES data within ±50 meV of the chemical potential and plotted as a function of $k_x$ and $k_y$. Data was obtained with 33 eV photons at T=20K in a half of the Brillouin zone and then reflected about e symmetry axis. Spacing between both sides of the Fermi surface was checked using wide angular scans such as those following in panels. The lines of high intensity (green, yellow and red) correspond to the Fermi surface. **b-e**, Intensity maps along $k_x$ taken at $k_y=0.53\pi/a$, $0.69\pi/a$, $0.84\pi/a$ and $\pi/a$, respectively (position of cuts are indicated by dotted lines in panel (**a**)). Experimental conditions were the same as in figure 1.

**Figure 3** Intensity maps and EDC spectra close to the chemical potential along selected cuts in the Brillouin zone using 33 eV photons at T=20K. **a-d**, Intensity maps for $k_y=0$, $0.52\ \pi/a$, $0.63\ \pi/a$ and $\pi/a$. **e-h** Energy distribution curves along the same cuts as in panels (**a-d**). **i** EDC spectra at $k_F$ for selected values of $k_y$. **j**

weight of the quasiparticle peak along the Fermi surface. The weight was estimated by calculating the area under the peak marked in red on the bottom EDC spectrum in panel (**j**). **k** EDC spectra for $k_y$=0.63 $\pi$/a (largest renormalization effects) at Fermi momentum below (blue) and above (red) the bulk superconducting temperature. **l** same as in (**k**), but for momentum slightly less than Fermi momentum.

**Figure 4** Momentum dependence of the superconducting gap in the quasi 1D conducting band. **a-f**, Symmetrized EDCs along selected cuts close to the Fermi momentum of the quasi 1D conducting band. In the symmetrization procedure the EDCs are reflected about the chemical potential, and then summed with non-reflected ones. This procedure removes the effect of the Fermi cut-off and allows one to easily identify even a small superconducting gap. The positions of the cuts are indicated by solid black lines in the next panel. **g**, Fermi surface of the planes (bonding: solid black lines and antibonding: dashed lines) and chains (red). A circle indicates the location of the maximum superconducting gap in the chain states. Magnitude of the superconducting gap at $k_F$ for different $k_y$ values. Data obtained under the same conditions as in figure 3.

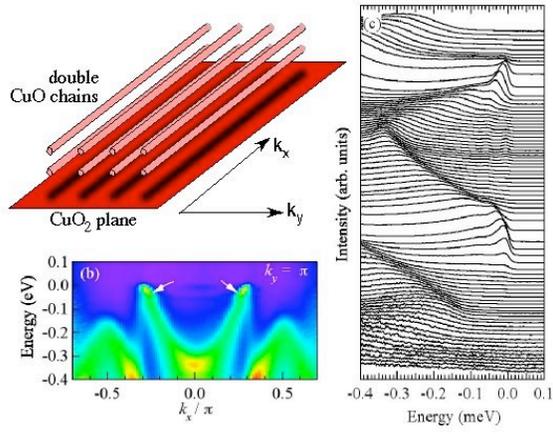

Figure 1

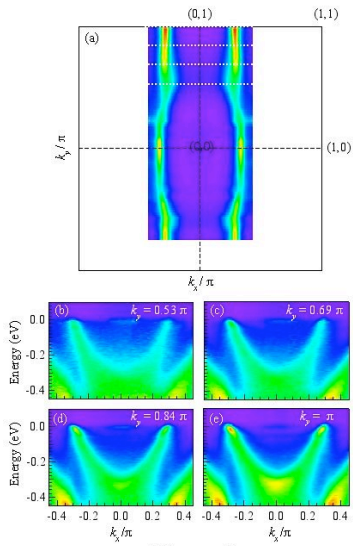

Figure 2

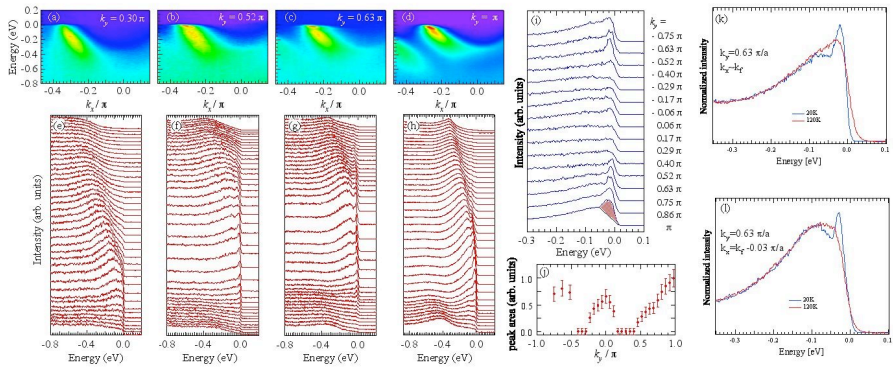

Figure 3

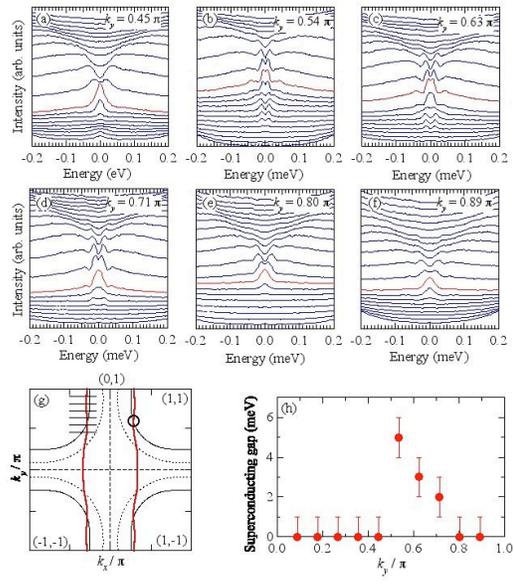

Figure 4